\documentclass[showpacs, pra,twocolumn,preprintnumbers ,amsmath, amssymb, superscriptaddress, aps]{revtex4-2}
\usepackage{color}
\usepackage{amsmath,amssymb}
\usepackage{pifont}% http://ctan.org/pkg/pifont
\usepackage{amssymb}  % More symbols
\usepackage{bbold}
\usepackage{float}
\usepackage{subfloat}
\usepackage[squaren]{SIunits}
\usepackage{comment}

\usepackage[caption=false]{subfig}
\usepackage{tikz}
\usepackage{makecell}
\usepackage{subfig}
\usepackage{pifont}   % Ding symbols
\usepackage{graphicx} % Include figure files
\graphicspath{{Figuress/}}
\usepackage{dcolumn}  % Align table columns on decimal point
\usepackage{bm}       % bold math
\usepackage{multirow} % Table functions
\usepackage{placeins}% For making floats not move around everywhere
\usepackage[colorlinks]{hyperref}
\usepackage{mathtools}
\usepackage{appendix}
\usepackage{orcidlink}

\begin{document}
	
	\title{Dynamics of quantum entanglement in two time-dependent coupled harmonic oscillators\\}  
	\date{\today}
	
	\author{Ayoub Ghaba \orcidlink{0009-0007-1263-1773}}
    \email{ayoub.ghaba-etu@etu.univh2c.ma}
	\affiliation{Laboratory of Mathematics and Physical Sciences Applied to Engineering Sciences, Faculty of Sciences and Techniques, Hassan II University, Mohammedia, Morocco}
	\author{Radouan Hab-arrih}
	\email{habarrih46@gmail.com}
	\affiliation{Laboratory of R\&D in Engineering Sciences, Faculty of Sciences and Techniques Al-Hoceima, Abdelmalek Essaadi University, Tetouan, Morocco.}
    \author{Elhoussine Atmani}
   \affiliation{Laboratory of Mathematics and Physical Sciences Applied to Engineering Sciences, Faculty of Sciences and Techniques, Hassan II University, Mohammedia, Morocco}
   \author{Abdallah Slaoui \orcidlink{0000-0002-5284-3240}}\email{abdallah.slaoui@um5s.net.ma}\affiliation{LPHE-MS, Faculty of Sciences, Mohammed V University in Rabat, Rabat, Morocco}\affiliation{CPM, Faculty of Sciences, Mohammed V University in Rabat, Rabat, Morocco.}


	\begin{abstract}
		We investigate the quantum entanglement dynamics of two coupled harmonic oscillators with a time-dependent interaction. Using the Lewis–Riesenfeld invariant method, we derive the exact analytical wave functions without any perturbative or adiabatic approximations and combine this with a phase-space analysis using the Wigner function to provide a complete description of the system's quantum state evolution. We obtain general expression form for the purity and the linear entropy $S_L=1-\mathcal{P}$ for arbitrary excitation numbers $(n,m)$, allows a systematic study of entanglement for a large class of quantum states. We show that the entanglement dynamics is very sensitive to the interplay between the detuning parameters $\theta$ and $\vartheta_2$, the frequency parameter $\beta_0$ and the coupling strength $\epsilon$: the increase of detuning takes the system from slow irregular oscillations to fast and regular periodic behavior, and the stronger coupling systematically enhances both the amplitude and the average value of the linear entropy. Most importantly, for the resonance case $\omega_1=\omega_2=1$ and strong couplings $\epsilon \approx 0.99$, the system shows robust undamped synchronized periodic oscillations of the linear entropy for all quantum states considered, indicating preserved quantum coherence without saturation. Our findings demonstrate that linear entropy is a sensitive and practical entanglement witness, and we establish explicit analytical relations between the coupling parameters of the system and its entanglement properties, which are directly relevant to quantum information processing and the control of quantum correlations in continuous-variable systems. 

	\end{abstract}

	\maketitle
	
\section{Introduction}
One of the most basic and versatile systems that has been extensively studied is the quantum harmonic oscillator, which accurately describes various classes of continuous variable (CV) systems \cite{Intro1,Intro2,1stchapter 9}. Continuous variable systems play a crucial role in current quantum communication and computing, with typical examples being quantized fields of electromagnetic in cavity QED \cite{Intro3,Intro4,Intro5}, optomechanical systems consisting of mechanical oscillators \cite{Intro6,Intro7}, and ion trap systems \cite{Intro8,Intro9,Intro10}. Due to its wide applicability, the dynamics of coupled oscillators have attracted considerable attention among physicists, especially when the coupling forces depend explicitly on time, enabling novel dynamical features not present in static models. The time dependence of oscillator systems has become a very useful tool indeed. It has been shown that within optomechanical systems, periodic modulation of the amplitude of driving makes it possible to obtain mechanical squeezing of great magnitudes without employing any form of feedback and squeezing of the light itself \cite{Intro11,Intro12,Intro13}. In Ref. \cite{Intro11}, it is found that one can significantly change the dynamical behavior of the entanglement by adding an extremely weak time modulation with a specific frequency to a constant coupling. enabling novel dynamical features not present in static models. \par

Entanglement, arguably the most important property of quantum mechanics, is of great significance for quantum information schemes and has been intensively studied in the context of coupled oscillator models \cite{1stchapter 9, Intro15, Intro16}. The vast majority of these works deal with stationary (time-independent) problems in which the entanglement structure is set by constant normal mode frequencies and coupling strength. In doing so, they obtain either stationary bounds for their correlations or a lack thereof. Such problems, despite being solvable by analytical methods, fall short of capturing the more complex quantum correlations present in a system where its parameters vary as a function of time. Time-dependent harmonic oscillator systems greatly broaden the possibilities: For optomechanical models, a modulation in the driving amplitude results in high mechanical squeezing without resorting to any feedback and squeezed light \cite{Intro11, Intro12, Intro13}, while nanomechanical and Dicke-driven models have time-modulated coupling resulting in dynamics inaccessible by stationary models \cite{Intro17, Intro14}. More surprising still, it was proven that the introduction of a small amount of time dependency to the coupling at a particular frequency results in drastic changes in the entanglement dynamics of the system \cite{Intro11}.\par

In this work, we investigate the problem of the dynamics of entangled states of a two-oscillator system with time-dependent coupling intensity by taking into account both the phase-space characteristics and entanglement features at once. In order to do so, we exploit the Lewis-Riesenfeld invariant technique, which helps to derive the exact solutions to the time-dependent Schrödinger equation and consequently derive analytical expressions for the wavefunctions without perturbative approximation. This approach enables us to define the Wigner function and explore the problems of phase-space fluctuations and uncertainty relations, thereby providing a deeper insight into the quantum behavior of the considered system in the phase-space. We use the linear entropy $S_L^{(n,m)} = 1-\mathcal{P}^{(n,m)}$ for all the states $(n,m)$, where $\mathcal{P}$ is the purity of a reduced subsystem's state, as the characteristic of bipartite entanglement. The results of our systematic analysis include three key findings. First, the detuning parameters $\theta$ and $\vartheta_{2}$ regulate the oscillatory dynamics of the linear entropy in such a way that, when their values are increased, slow and chaotic oscillations turn into faster and more coherent oscillations, with the parameter $\theta$ being responsible for controlling the oscillation amplitude. Second, the frequency parameter $\beta_{0}$ enables very accurate regulation of the entropy oscillations, while the coupling parameter $\epsilon$ influences the results more strongly, as it not only increases the oscillation amplitude but also the average value of the entropy, which demonstrates a higher degree of entanglement between the oscillators. Third, at strong coupling ($\epsilon \approx 0.99$), resonance frequency ($\omega_{1}=\omega_{2}=1$) causes undamped and synchronized oscillations of linear entropy for all quantum states without reaching the saturation point. Entanglement dynamics are highly sensitive to the interplay between detuning, frequency, and coupling parameters; even small changes can cause drastic changes in the overall dynamics behavior. These findings are consistent with, and extend, the results of Galve et al. \cite{Intro18}, Roque and Roversi \cite{Intro19}, and Bastidas et al. \cite{Intro14}, who collectively established that the dynamical stability of the normal modes governs the entanglement behavior in time-dependent coupled oscillator systems. Our work reinforces and enriches this picture analytically through the Lewis–Riesenfeld invariant method, providing explicit and systematic control over the entanglement dynamics via the parameter set $(\theta, \vartheta_2, \beta_0, \epsilon)$.\par

The remainder of this paper is organized as follows. In Sec. (\ref{sec2}), we present the physical model of the two coupled time-dependent harmonic oscillators and derive the exact wave functions via the Lewis–Riesenfeld invariant method. In Sec. (\ref{sec3}), we construct the Wigner function, analyze the phase space fluctuations, and examine the Heisenberg uncertainty relations. In Sec. (\ref{sec4}), we measure the bipartite entanglement through the linear entropy.  In Sec. (\ref{sec4}), we provide a detailed analysis of its dependence on the detuning, frequency, and coupling parameters, including the strong coupling resonance regime for different initial quantum states. Finally, in Sec. (\ref{sec6}), We provide a summary of our main discoveries.

\section{The Hamiltonian and  Wave Function via Lewis-Riesenfeld Method \label{sec2}}
We consider a system of two coupled harmonic oscillators governed by the following time-dependent Hamiltonian.
\begin{eqnarray}
\hat{\mathbb{H}}&=&\sum _{i=1}^{2}\left( \frac{\hat{p}_i^2}{2M(t)}+\frac{1}{2}M(t)\omega_i(t)^2 \hat{x}_i^2 \right)-\epsilon(t) \hat{x_1}\hat{x_2} \label{Eq1}
\end{eqnarray}
From now on, for simplicity, we consider $\hbar=1$ \cite{1stchapter 1}. The momentum and position operators fulfill the commutation relations. $[\hat{x_1},\hat{p_1}]=[\hat{x_2},\hat{p_2}]=\textbf{i}$ and $[\hat{x}_1,\hat{x}_2]=[\hat{p}_1,\hat{p}_2]=0$. 
The Hamiltonian \eqref{Eq1} can be instantly diagonalized by rotating the variables 
\begin{equation}
    \binom{\hat{X_1}}{\hat{X_2}} =  \mathbb{R}(\theta)\binom{\hat{x}_1}{\hat{x}_2}, \quad 
   \binom{\hat{P}_1}{\hat{P}_2} =  \mathbb{R}(\theta)\binom{\hat{p}_1}{\hat{p}_2}
\end{equation}
Where the rotation matrix, $\mathbb{R}$, is defined as
\begin{equation}
    \mathbb{R}(\theta) = \begin{pmatrix} 
    C_\theta & S_\theta \\ 
    -S_\theta  & C_\theta  
\end{pmatrix}
\end{equation}
The shorthand notation $(S_\theta, C_\theta) \equiv (\sin \theta, \cos \theta)$ is used here and will be used for the rest of the manuscript.\\
Our new coordinates are given as
\begin{align}
&\label{eq5}	\hat{X}_1= \hat{x}_1 C_\theta +\hat{x}_2 S_\theta , \quad \hat{X}_2= -\hat{x}_1 S_\theta +\hat{x}_2 C_\theta\\
&\label{eq6}	\hat{P}_1= \hat{p}_1 C_\theta +\hat{p}_2 S_\theta, \quad \hat{P}_2= -\hat{p}_1 S_\theta+\hat{p}_2 C_\theta 
\end{align}
The Hamiltonian \eqref{Eq1}, expressed in terms of the new variables 
\begin{equation}\label{eq3}
\hat{\mathbb{H}}_d=\frac{1}{2M(t)} \left(\hat{P}_1^2+\hat{P}_2^2\right)+\frac{1}{2}\vartheta^{2}_{1}(t)\hat{X}_1^2
+\frac{1}{2}\vartheta^{2}_{2}(t)\hat{X}_2^2
\end{equation}
Where $\theta$, the rotation angle, is 
\begin{eqnarray}
\theta=\frac{1}{2}\arctan\left(\frac{2\epsilon(t)}{M(t)\big(\omega_1^2(t)-\omega_{2}^2(t)\big)}\right)=\sf constant.
\end{eqnarray}
The mixing $\theta$ is constant over time, meaning that $(\dot{\theta}=0)$ \cite{Redoaun1}. The normal frequencies, which are written as 
\begin{align}
\vartheta_{1,2}^2&=\pm\frac{1}{2}\sqrt{M^2(t)\left[\omega_1^2(t)-\omega_2^2(t)\right]^2+4\epsilon^2(t)}\notag\\&+\frac{M(t)\big(\omega_1^2(t)+\omega_2^2(t)\big)}{2}\notag\\&=\omega_{1,2}^2(t)\pm\frac{\epsilon(t)}{M(t)} \tan\theta \label{Eq5}
\end{align}
The LR (Lewis-Riesenfeld) invariant satisfies the  \cite{1stchapter 2,1stchapter 12,1stch9}
\begin{equation}
      \frac{dI(t)}{dt} = \frac{\partial I}{\partial t} + \frac{1}{i\hbar}[I, \hat{\mathbb{H}}_d] = 0, \quad I^\dagger = I
\end{equation}
Which can generally be expressed in the following form \cite{1stchapter 3}
\begin{equation}
    \hat{I}(t) =\sum_{i=1}^{2} \mathbb{g}_{-i}(t) \frac{\hat{P}_i^2}{2} + \mathbb{g}_{0i}(t) \frac{\hat{P}_i\hat{X}_i + \hat{X}_i\hat{P}_i}{2} + \mathbb{g}_{+i}(t) \frac{\hat{X}_i^2}{2}
\end{equation}
where $\mathbb{g}_{j}(t) (j = -, 0, +)$ satisfies the linear system of differential equations and can be expressed using the parameters of the system as \cite{1stchapter 3,1stchapter 4}
\begin{align}
\dot{\mathbb{g}}_{-i}(t) &= -\frac{2}{M(t)} g_{0i}(t)\\
\dot{\mathbb{g}}_{0i}(t) &= M(t)\vartheta_i^{2}(t)\mathbb{g}_{-i}(t) - \frac{1}{M(t)}g_{+i}(t)\\
\dot{\mathbb{g}}_{+i}(t) &= 2M(t)\vartheta_i^{2}(t)g_{0i}(t).
\end{align}
The system's general solution is represented by the classical equation $g–i(t)$ in the following form
\begin{align}
    \mathbb{g}_{-1}(t) &= \eta_1(t)\, \mathcal{F}_1^2(t) 
          + \beta_1(t)\, \mathcal{F}_1(t) \mathcal{G}_1(t) 
          + \gamma_1(t)\, \mathcal{G}_1^2(t), \label{equation15}\\
    \mathbb{g}_{-2}(t) &= \eta_2(t)\, \mathcal{F}_2^2(t) 
          + \beta_2(t)\, \mathcal{F}_2(t) \mathcal{G}_2(t) 
          + \gamma_2(t)\, \mathcal{G}_2^2(t) \label{eq16}
\end{align}
where $\eta_i(t),\,\beta_i(t),\,\gamma_i(t)$ are arbitrary constants. Additionally, $\mathcal{F}_i,\,\mathcal{G}_i$ are two linearly independent solutions of the classical equations of motion
\begin{align}
     &\ddot{\mathcal{F}_i}(t) + \frac{\dot{M}(t)}{M(t)} \, \dot{\mathcal{F}_i}(t) + \vartheta_i^2(t) \mathcal{F}_i(t) = 0, \\
     &\ddot{\mathcal{G}_i}(t) + \frac{\dot{M}(t)}{M(t)} \, \dot{\mathcal{G}_i}(t) + \vartheta_i^2(t) \mathcal{G}_i(t) = 0, 
\quad i = 1,2
\end{align}
Under the condition of being $g_{-}(t)$ always positive because of the creation and annihilation operators, it shouldn't be singular \cite{1stchapter 3}, the LR invariant can be written as 
\begin{equation} \label{equation19}
\hat{I}(t)=\hat{I}_1(t)+\hat{I}_2(t)=\sum_{i=1}^{2}\frac{1}{2} \Bar{P}_i^2+\frac{1}{2}\Omega_i^{2}(t)\Bar{X}_i^{2}.
\end{equation}
Here, the frequency $\Omega_i$, derived from the diagonalized LR invariant  
\begin{equation}
    \Omega_i(t)=\bigg[ \mathbb{g}_{+i}(t)\mathbb{g}_{-i}(t)-\mathbb{g}_{0i}^2(t)\bigg]^{1/2}. \label{eq20}
\end{equation}
So, the new canonical variables $\bar{P}_i(t)$ and $\bar{X}_i(t)$ come from two consecutive time dependent unitary transformations
\begin{align}
    &U_{1}(t)=\exp\bigg(\textbf{i}\frac{\mathbb{g}_{0i}(t)}{2\mathbb{g}_{-i}(t)}\hat{X}_i\bigg),\notag\\&
    U_{2}(t)=\exp\bigg(\frac{\textbf{i}}{4}\big(\mathbb{P}_i\mathbb{X}_i+\mathbb{X}_i\mathbb{P}_i\big)\ ln\big(\mathbb{g}_{-i}(t)\big)\bigg),
\end{align}
with 
\begin{equation}\label{eq18}
    \mathbb{X}_i= U_{1}^{-1} \hat{X}_i U_{1}=\hat{X}_i,  \quad \mathbb{P}_i= U_{1}^{-1} \hat{P}_i U_{1}=\hat{P}_i+\frac{g_{0i}}{\mathbb{g}_{-i}}\hat{X}_i
\end{equation}
and 
\begin{equation}\label{eq19}
    \Bar{X}_i= U_{2}^{-1} \mathbb{X}_i U_{2}=\frac{1}{\sqrt{\mathbb{g}_{-i}}}\mathbb{X}_i,  \quad \bar{P}_i= U_{2}^{-1} \mathbb{P}_i U_{2}=\sqrt{\mathbb{g}_{-i}}\mathbb{P}_i
\end{equation}
We have $[\bar{X}_i, \bar{P}_i] = [\hat{x}_i, \hat{p}_i]=\textbf{i}$, indicating that the commutation relations are consistent across both coordinate systems. The associated eigenfunctions defined as
\begin{widetext}
\begin{align}\label{eq21}
	\Psi_{(n,m)}(\hat{X}_1,\hat{X}_2,t)&= \Psi_n(\hat{X}_1,t)\otimes \Psi_{m}(\hat{X}_2,t)\notag\\ \notag
	&=\frac{1}{\sqrt{2^{n+m}n!m!}}\left(\frac{\Omega_1\Omega_2}{\pi^2\mathbb{g}_{-1}(t)\mathbb{g}_{-2}(t)}\right)^{\frac{1}{4}}
    e^{-\textbf{i}\frac{\mathbb{g}_{01}(t)}{2\mathbb{g}_{-1}(t)}\hat{X}_1^2}e^{-\textbf{i}\frac{\mathbb{g}_{02}(t)}{2\mathbb{g}_{-2}(t)}\hat{X}_2^2} e^{-\textbf{i}\big(\alpha_n(t)+\alpha_m(t)\big)}e^{-\frac{\Omega_1}{2\mathbb{g}_{-1}}\hat{X}_1^2}e^{-\frac{\Omega_2}{2\mathbb{g}_{-2}}\hat{X}_2^2}\notag\\
    & \phantom{==} \times H_n\bigg(\sqrt{\frac{\Omega_1}{\mathbb{g}_{-1}(t)}}\hat{X}_1\bigg)H_m\bigg(\sqrt{\frac{\Omega_2}{\mathbb{g}_{-2}(t)}}\hat{X}_2\bigg),
\end{align} 
with 
$H_n(x)$ are for Hermite polynomials, and the parametrs $ \alpha_n(t)$ and $\alpha_m(t)$ are defined as result
\begin{equation}
    \alpha_n(t)= \left(n + \frac{1}{2}\right)\int_0^t \frac{\Omega_1(t_1)}{M(t)\mathbb{g}_{-1}(t_1)} \, dt_1
 \hspace{2cm} \alpha_m(t)=\left(m + \frac{1}{2}\right)\int_0^t \frac{\Omega_2(t_2)}{M(t)\mathbb{g}_{-2}(t_2)}\, dt_{2}.
\end{equation}
In terms of the original coordinates $(\hat{x}_1,\hat{x}_2)$, the wave function in Eq.(\ref{eq21}) becomes
\begin{align}
	\Psi_{(n,m)}(\hat{x}_1,\hat{x}_2,t)&= \Psi_n(\hat{x}_1,t)\otimes \Psi_{m}(\hat{x}_2,t)\notag\\ \notag
	&=\frac{1}{\sqrt{2^{n+m}n!m!}}\left(\frac{\Omega_1\Omega_2}{\pi^2\mathbb{g}_{-1}(t)\mathbb{g}_{-2}(t)}\right)^{\frac{1}{4}}
    e^{-\textbf{i}\frac{\mathbb{g}_{01}(t)}{2\mathbb{g}_{-1}(t)}\big[\hat{x}_1 C_\theta (t) +\hat{x}_2 S_\theta(t)\big]^2}e^{-\textbf{i}\frac{\mathbb{g}_{02}(t)}{2\mathbb{g}_{-2}(t)}\big[-\hat{x}_1 S_\theta(t) +\hat{x}_2 C_\theta(t)\big]^2}\\
    &\phantom{==} e^{-\textbf{i}\big(\alpha_n(t)+\alpha_m(t)\big)}e^{-\frac{\Omega_1}{2\mathbb{g}_{-1}}\big[\hat{x}_1 C_\theta (t) +\hat{x}_2 S_\theta(t)\big]^2}e^{-\frac{\Omega_2}{2\mathbb{g}_{-2}}\big[-\hat{x}_1 S_\theta(t) +\hat{x}_2 C_\theta(t)\big]^2}\notag\\
    &\phantom{==} \times H_n\bigg(\sqrt{\frac{\Omega_1}{\mathbb{g}_{-1}(t)}}\big[\hat{x}_1 C_\theta (t) +\hat{x}_2 S_\theta(t)\big]\bigg)H_m\bigg(\sqrt{\frac{\Omega_2}{\mathbb{g}_{-2}(t)}}\big[-\hat{x}_1 S_\theta(t) +\hat{x}_2 C_\theta(t)\big]\bigg)
\end{align}
\end{widetext}

At initial time $t_{0}$, we have \cite{1stchapter 3}
\begin{align}\label{eq29}
&\mathbb{g}_{-i}(t_0) = \frac{1}{M(t_0)}, \hspace{1cm}
\mathbb{g}_{0i}(t_0) = 0, \notag\\&
\mathbb{g}_{+i}(t_0) = M(t_0) \vartheta_i^2(t_0), \hspace{1cm} \Omega_i(t_0)=\vartheta_i(t_0).
\end{align}
By adjusting these parameters, we achieved $H(t_0) = I(t_0)$. In this case, the Hamiltonian falls back to that of a time-independent oscillator.
\section{Wigner function, phase space fluctuations and uncertainty relations\label{sec3}}

\subsection{Wigner function}
Among numerous representations of the quantum state, the Wigner function provides a particularly appealing framework for describing quantum phenomena through a classical-like phase space formalism \cite{ref7,ref8}. This approach yields far more information about the system than other quantum mechanical approaches \cite{wigner1} and offers complete information on the state of the system \cite{ref8}. Specifically, the Wigner function corresponding to the diagonalized Hamiltonian in Eq.(\ref{equation19}) exhibits a separable structure
\begin{eqnarray}\label{10}
W_{(n,m)}(\bar{X}_1,\bar{P}_1;\bar{X}_2,\bar{P}_2;t)= W_{n}(\bar{X}_1,\bar{P}_1;t)\times W_m(\bar{X}_2,\bar{P}_2;t),
\end{eqnarray}
where the functions involved are \cite{1stchapter 7}
\begin{align}
W_n(\bar{X}_1,\bar{P}_1,t)&=\frac{1}{\pi} \int d\mathbb{\bar{X}_1}\ \Psi_{n}^{\ast}(\bar{X}_1+\mathbb{\bar{X}_1})\ \Psi_{n}(\bar{X}_1-\mathbb{\bar{X}_1})\ e^{2i\Bar{P}_1\mathbb{\bar{X}_1}}\notag\\\notag
&=\frac{(-1)^n}{\pi}e^{-\frac{2}{\Omega_1}\hat{I}_1(\bar{X}_1,\bar{P}_1,t)}\mathcal{L}_{n}\left[\frac{4}{\Omega_1}\hat{I}_1(\bar{X}_1,\bar{P}_1,t)\right] \\
&=\frac{(-1)^n}{\pi}e^{-\frac{1}{\Omega_1}\left({\Omega_1^{2}}\bar{X_1}^2+\bar{P}_1^{2}\right)}\mathcal{L}_{n}\left[\frac{2}{\Omega_1}\left(\Omega_1^{2}\bar{X}_1^2+\bar{P}_1^{2}\right)\right]
\end{align}
and
\begin{align}
W_m(\bar{X}_2,\bar{P}_2,t)&=\frac{1}{\pi} \int d\mathbb{\bar{X}_2}\ \Psi_{m}^{\ast}(\bar{X}_2+\mathbb{\bar{X}_2})\ \Psi_{m}(\bar{X}_2-\mathbb{\bar{X}_2})\ e^{2i\Bar{P}_2\mathbb{\bar{X}_2}}\\\notag
&=\frac{(-1)^m}{\pi}e^{-\frac{2}{\Omega_2}\hat{I}_2(\bar{X}_2,\bar{P}_2,t)}\mathcal{L}_{m}\left[\frac{4}{\Omega_2}\hat{I}_2(\bar{X}_2,\bar{P}_2,t)\right] \notag\\
&=\frac{(-1)^m}{\pi}e^{-\frac{1}{\Omega_2}\left({\Omega_2^{2}}\bar{X}_2^2+\bar{P}_2^{2}\right)}\mathcal{L}_{m}\left[\frac{2}{\Omega_2}\left(\Omega_2^{2}\bar{X}_2^2+\bar{P}_2^{2}\right)\right]
\end{align}
The global Wigner function can be written in terms of the variables $(\bar{X}_i,\bar{P}_i)$ as
\begin{widetext}
\begin{equation}\label{eq31}
W_{(n,m)}(\bar{X}_1,\bar{P}_1;\bar{X}_2,\bar{P}_2;t)= \frac{(-1)^{n+m}}{\pi^2}e^{-\frac{1}{\Omega_1}\left({\Omega_1^{2}}\bar{X_1}^2+\bar{P}_1^{2}\right)-\frac{1}{\Omega_2}\left({\Omega_2^{2}}\bar{X}_2^2+\bar{P}_2^{2}\right)}\mathcal{L}_{n}\left[\frac{2}{\Omega_1}\left(\Omega_1^{2}\bar{X}_1^2+\bar{P}_1^{2}\right)\right]\mathcal{L}_{m}\left[\frac{2}{\Omega_2}\left(\Omega_2^{2}\bar{X}_2^2+\bar{P}_2^{2}\right)\right]
\end{equation}
\end{widetext}
and $\mathcal{L}_{n}(x)$ are for Laguerre polynomials\cite{1stchapter 6}. Using the variables~(\ref{eq5}),~(\ref{eq6}),~(\ref{eq18}) and (\ref{eq19}),  we map the variables $(\bar{X}_i,\bar{P}_i)$ in terms of the original coordinates
\begin{align}
     &\bar{X}_1=\frac{1}{\sqrt{\mathbb{g}_{-1}}}\big(\hat{x}_1 C_\theta (t) +\hat{x}_2 S_\theta(t) \big), \notag\\& \bar{P}_1=\sqrt{\mathbb{g}_{-1}}\bigg[\big(\hat{p}_1 C_\theta (t) +\hat{p}_2 S_\theta(t) \big)+\frac{\mathbb{g}_{01}}{\mathbb{g}_{-1}}\big(\hat{x}_1 C_\theta (t) +\hat{x}_2 S_\theta(t) \big)\bigg],\label{eq35}
\end{align}
and
\begin{align}
     &\bar{X}_2=\frac{1}{\sqrt{\mathbb{g}_{-2}}}\big( -\hat{x}_1 S_\theta(t) +\hat{x}_2 C_\theta(t)  \big), \notag\\& \bar{P}_2=\sqrt{\mathbb{g}_{-2}}\bigg[\big(-\hat{p}_1 S_\theta(t) +\hat{p}_2 C_\theta(t)  \big)+\frac{\mathbb{g}_{02}}{\mathbb{g}_{-2}}\big(-\hat{x}_1 S_\theta(t) +\hat{x}_2 C_\theta(t)  \big)\bigg],\label{eq36}
\end{align}
and we can extract the original variables as function of the canonical ones, which is written as 
\begin{align}
     &\hat{x}_1= \sqrt{\mathbb{g}_{-1}} C_{\theta}\bar{X}_1-\sqrt{\mathbb{g}_{-2}} S_{\theta}\bar{X}_{2}, \notag\\& \hat{p}_1 =C_{\theta}\bigg(\frac{\bar{P} _1 }{\sqrt{\mathbb{g}_{-1}}}-\frac{g\mathbb{g}_{01}}{\mathbb{g}_{-1}} \bar{X}_1\bigg)-S_{\theta}\bigg(\frac{\bar{P}_2}{\sqrt{\mathbb{g}_{-2}}}-\frac{\mathbb{g}_{02}}{\mathbb{g}_{-2}}\bar{X}_2\bigg), \label{eq38}
\end{align}
and
\begin{align}
   &\hat{x}_2= \sqrt{\mathbb{g}_{-1}} S_{\theta}\bar{X}_1+
    \sqrt{\mathbb{g}_{-2}} C_{\theta}\bar{X}_2 , \notag\\& \hat{p}_2 =S_{\theta} \left(\frac{\bar{P}_1}{\sqrt{\mathbb{g}_{-1}}}-\frac{\mathbb{g}_{01}}{\mathbb{g}_{-1}}\bar{X}_1\right)+C_{\theta} \left(\frac{\bar{P} _2}{\sqrt{\mathbb{g}_{-2}}}-\frac{\mathbb{g}_{02}}{\mathbb{g}_{-2}}\bar{X}_2\right). \label{eq39}
\end{align}
The Wigner function will play a crucial role in analyzing the entanglement and steering our system.
\subsection{Phase space fluctuations}
Several quantum phenomena, including entanglement, quantum tunneling, and quantum steering, depend on the quantum fluctuations. To comprehend the behavior of quantum systems at the microscopic level, it is crucial to Characterizing and understanding these quantum fluctuations. Here we examine the quantum quantum fluctuation using the canonical coordinates $(\Bar{X}_i,\Bar{P}_i)$, and the original variable $(\hat{x}_i,\hat{p}_i) $

\subsubsection{Expectation value using the original variables }
The expectation values as a function of the original coordinates $(\hat{x}_i,\hat{p}_i)$ are given as
\begin{eqnarray}
\langle\mathcal{T}\rangle=\int_{\mathbb{R}^4}d\hat{x}_1 d\hat{x}_2 d\hat{p}_1 d\hat{p}_2\
\mathcal{T} \ W_{(n,m)}(\hat{x}_1,\hat{x}_2,\hat{p}_1,\hat{p}_1,t) 
\end{eqnarray}
As a consequence, we display the following expectation values for the positions and moments
\begin{align}
&\langle \hat{x}_1^2\rangle =\frac{(1+2n)\mathbb{g}_{-1}}{2\Omega_1 (\mu_{\theta}^2+1)}+\frac{(1+2m)\mu_{\theta}^2\mathbb{g}_{-2}}{2\Omega_2 (\mu_{\theta}^2+1)},\notag\\&  \langle \hat{x}_2^2\rangle=\frac{(1+2n)\mathbb{g}_{-1}\mu_{\theta}^2}{
2\Omega_1 (\mu_{\theta}^2+1)}+\frac{(1+2m)\mathbb{g}_{-2}}{2\Omega_2 (\mu_{\theta}^2+1)}, \notag\\& \langle \hat{p}_1^2\rangle =\frac{(2 n+1)\mathbb{g}_{-1} \mathcal{K}_1}{2\Omega _1 \left(\mu _{\theta }^2+1\right)}+\frac{(2 m+1) \mu _{\theta }^2\mathbb{g}_{-2}\mathcal{K}_2}{2\Omega _2 \left(\mu _{\theta }^2+1\right)},\notag\\&  \langle \hat{p}_2^2\rangle=\frac{(1+2n)\mathbb{g}_{-1}\mu_{\theta}^2\mathcal{K}_1}{
2\Omega_1 (\mu_{\theta}^2+1)}+\frac{(1+2m)\mathbb{g}_{-2}\mathcal{K}_2}{2\Omega_2 (\mu_{\theta}^2+1)}.
\end{align}
With the substitution, $\mathcal{K}_i$ can be expressed as 
\begin{align}
    \mathcal{K}_1= \frac{\mathbb{g}_{01}^2+\Omega _1^2 \mathbb{g}_{-1}}{\mathbb{g}_{-1}^3}&,\qquad  \mathcal{K}_2=\frac{\mathbb{g}_{02}^2+\Omega _2^2\mathbb{g}_{-2}}{\mathbb{g}_{-2}^3}\label{eq97}
\end{align}
At the initial time $(t_{0})$ in Eq.(\ref{eq29}), with a unit mass $(M(t_0)=1)$, the results align with those documented in the literature \cite{1stchapter 9}. The substitution in Eq.~(\ref{eq97}) reduce to:
\begin{align}
     \mathcal{K}_1= \vartheta^2_{1}&,\qquad  \mathcal{K}_2=\vartheta^2_{2}
\end{align}
These findings will be used in the next part to get further insight into the current system.
\subsection{Analysis of Heisenberg Uncertainties}
The following is a presentation of the following phase space areas that are associated with the two oscillators as a function of the coordinates $(\hat{x}_i,\hat{p}_i)$, the uncertainty relations for various states take the form
\begin{widetext}
\begin{align}
[\mathfrak{A}_1(n,m)]^2&=\left[\Delta \hat{x}_1\Delta \hat{p}_1 (n,m)\right]^2\notag\\
&=\left(\frac{(2 n+1) \mathbb{g}_{-1}}{2 \Omega _1 \left(\mu _{\theta }^2+1\right)}+\frac{(2 m+1)\mathbb{g}_{-2} \mu _{\theta }^2}{2 \Omega _2 \left(\mu _{\theta }^2+1\right)}\right) \left(\frac{(2 n+1) \left(\mathbb{g}_{01}^2+\Omega _1^2 \mathbb{g}_{-1}\right)}{2 \Omega _1 \mathbb{g}_{-1}^2 \left(\mu _{\theta }^2+1\right)}+\frac{(2 m+1) \mu _{\theta }^2 \left(\mathbb{g}_{02}^2+\Omega _2^2\mathbb{g}_{-2}\right)}{2 \Omega _2\mathbb{g}_{-2}^2 \left(\mu _{\theta }^2+1\right)}\right)
\\
[\mathfrak{A}_2(n,m)]^2&=
\left[\Delta \hat{x}_2\Delta \hat{p}_2 (n,m)\right]^2\notag\\
&=\left(\frac{(2 n+1) \mathbb{g}_{-1} \mu _{\theta }^2}{2 \Omega _1 \left(\mu _{\theta }^2+1\right)}+\frac{(2 m+1)\mathbb{g}_{-2}}{2 \Omega _2 \left(\mu _{\theta }^2+1\right)}\right) \left(\frac{(2 n+1) \mu _{\theta }^2 \left(\mathbb{g}_{01}^2+\Omega _1^2 \mathbb{g}_{-1}\right)}{2 \Omega _1 \mathbb{g}_{-1}^2 \left(\mu _{\theta }^2+1\right)}+\frac{(2 m+1) \left(\mathbb{g}_{02}^2+\Omega _2^2\mathbb{g}_{-2}\right)}{2 \Omega _2\mathbb{g}_{-2}^2 \left(\mu _{\theta }^2+1\right)}\right)
\end{align}

where $\Delta \mathcal{O} = \left( \langle \mathcal{O}^2 \rangle - \langle \mathcal{O} \rangle^2 \right)^{1/2}$ for any quantity $\mathcal{O}$ \cite{1stchapter 8}. Thus, the difference is given as 
\begin{align}
    [\mathfrak{A}_1(n,m)]^2-[\mathfrak{A}_2(n,m)]^2=\frac{(2 m+1)^2 g_{02}^2 \left(\mu _{\theta }^2-1\right)}{4 \Omega _2^2\mathbb{g}_{-2} \left(\mu _{\theta }^2+1\right)}-\frac{(2 n+1)^2 g_{01}^2 \left(\mu _{\theta }^2-1\right)}{4 \Omega _1^2 \mathbb{g}_{-1} \left(\mu _{\theta }^2+1\right)}-\frac{\left(\mu _{\theta }^2-1\right) (n-m) (m+n+1)}{\mu _{\theta }^2+1}
\end{align}
\end{widetext}
When the oscillators are in resonance ($\mu_{\theta}=1$), this results in equal areas in phase space. 
\begin{eqnarray}
\mathfrak{A}_1(n,m)=\mathfrak{A}_2(n,m)\label{resphase}
\end{eqnarray}
In the case when the oscillators are decoupled ($\mu_{\theta}=0$), the uncertainty relations take the following form 
\begin{align}
    [\mathfrak{A}_1(n,m)]^2=\frac{(2 n+1)^2 \left(g_{01}^2+\Omega _1^2 \mathbb{g}_{-1}\right)}{4 \Omega _1^2 \mathbb{g}_{-1}},
\end{align}
\begin{align}
[\mathfrak{A}_2(n,m)]^2=\frac{(2 m+1)^2 \left(g_{02}^2+\Omega _2^2\mathbb{g}_{-2}\right)}{4 \Omega _2^2\mathbb{g}_{-2}}
\end{align}

We now consider the scenario in which the frequencies of the two oscillators approach equality, i.e., $\Omega_1 \sim\Omega_2 \sim \Omega$, and for the symmetric state where $n=m$, resulting in the following result
\begin{align}
&[\mathfrak{A}_1(n,m)]^2-[\mathfrak{A}_2(n,m)]^2= \notag \\
&\qquad \frac{(2 n+1)^2 \left(\mu _{\theta }^2-1\right) \left(g_{02}^2 \mathbb{g}_{-1}-g_{01}^2\mathbb{g}_{-2}\right)}{4 \Omega ^2 \mathbb{g}_{-1}\mathbb{g}_{-2} \left(\mu _{\theta }^2+1\right)}
\end{align}
We notice even for the symmetric stats, and under the same frequency, the oscillators don't occupy the same surface in the phase space, and depend on the physical parameters of the system.
\section{Quantum entanglement \label{sec4}}
\subsection{Linear entropy}
The linear entropy is a fundamental measure in quantum information theory that quantifies the degree of mixedness of a quantum system \cite{entropy1}. Measurements of purity, or equivalently of linear entropy , remain the best candidates for a direct estimation of CV entanglement in realistic experiments \cite{entropy2}. The linear entropy can be expressed as \cite{1stchapter 8,1stchapter 11,entropy2}
\begin{equation}
S_L^{(n,m)}=1-\mathcal{P}^{(n,m)}
\end{equation}
Where $\mathcal{P}$ is the purity of the system\\
A pure quantum state has zero linear entropy ($S_L = 0$) and unit purity ($\mathcal{P} = 1$), while a maximally mixed state approaches maximum linear entropy ($S_L \to 1$) and zero purity ($\mathcal{P} \to 0$).
\subsection{Analysis of System Purity}
We start with the purity of the global state, which can be expressed as follows
\begin{equation}
4\pi^2\int_{\mathbb{R}^4}d\hat{x}_1\ d\hat{x}_2\ d\hat{p}_1\ d\hat{p}_2\ W^{2}_{(n,m)}(\hat{x}_1,\hat{p}_1,\hat{x}_2,\hat{p}_2)=1 \label{eq61}
\end{equation}
Where $W_{(n,m)}(\hat{x}_1,\hat{p}_1,\hat{x}_2,\hat{p}_2)$ is given by (\ref{eq31}) as a function of the original coordinates in Eqs. (\ref{eq35}), and (\ref{eq36})\\
Accordingly, entanglement can be assessed by calculating the marginal purities, specifically by evaluating one of the following purities
\begin{align}
	\mathcal{P}_{\hat{x}_1}^{(n,m)} &= 2\pi \int_{\mathbb{R}^2} dx \, dp \, W_{(n,m)}^2(\hat{x}_1,\hat{p}_1) \\
	\mathcal{P}_{\hat{x}_2}^{(n,m)} &= 2\pi \int_{\mathbb{R}^2} dy \, dq \, W_{(n,m)}^2(\hat{x}_2,\hat{p}_2)
\end{align}
where \( W_{(n,m)}(\hat{x}_1,\hat{p}_1) \) and \( W_{(n,m)}(\hat{x}_2,\hat{p}_2) \) are the marginal Wigner functions defined by
\begin{align}
	W_{(n,m)}(\hat{x}_1,\hat{p}_1) &= \int_{\mathbb{R}^2} dy \, dq \, W_{(n,m)}(\hat{x}_1,\hat{p}_1;\hat{x}_2,\hat{p}_2) \\
	W_{(n,m)}(\hat{x}_2,\hat{p}_2) &= \int_{\mathbb{R}^2} dx \, dp \, W_{(n,m)}(\hat{x}_1,\hat{p}_1;\hat{x}_2,\hat{p}_2)
\end{align}
This will allow us to calculate the purity in phase space for each oscillator and thus quantify the entanglement of the system. Additionally, the global state is a pure state (\ref{eq61}), and thus $\mathcal{P}_{\hat{x}_1}^{(n,m)}=\mathcal{P}_{\hat{x}_2}^{(n,m)}$. It is interesting to note that the calculation of the above integrals is not trivial. In order to proceed, we will follow a scheme based on the Rodrigues formula for Laguerre polynomials. \cite{1stchapter 6,1stchapter 10}
\begin{eqnarray}
	\mathcal{L}_n(x) &=& \frac{1}{n!} \frac{d^n}{du^n} \left( \frac{e^{-\frac{xu}{1-u}}}{1-u} \right) \Bigg\vert_{u=0}. \label{laguerre}
\end{eqnarray}
By integrating over \(\hat{x}_2\) and \(\hat{p}_2\) and using (\ref{laguerre}), we get the following purity expression
\begin{equation}
	\mathcal{P}^{(n,m)} = \frac{(\mu _{\theta }^2+1)}{(n! m!)^2} \frac{d^n}{du^n} \frac{d^m}{dv^m} \frac{d^n}{da^n} \frac{d^m}{db^m}	
	\left[\sqrt{\frac{1}{(\mathcal{A}+\mathcal{B})^2+\mathcal{A} \mathcal{B} \mathcal{Q}}}\right]_{u,s,v,w=0} \label{purity}
\end{equation}
Where $\mu_{\theta}=\tan(\theta)$\\
We set up the following functions 
\begin{align}
    &\mathcal{A}(u,v,a,b)=(b+1) (v+1) (a u-1), \notag\\&\mathcal{B}(u,v,a,b)= (a+1) (u+1) (b v-1) \mu _{\theta }^{2}, \label{eq58}
\end{align}
and 
\begin{equation}
        \mathcal{Q}(t)=\frac{\left(g_{02} \mathbb{g}_{-1}-g_{01}\mathbb{g}_{-2}\right)^2+\left(\Omega _1\mathbb{g}_{-2}-\Omega _2 \mathbb{g}_{-1}\right)^2}{\Omega _1 \Omega _2 \mathbb{g}_{-1}\mathbb{g}_{-2}} \label{eq65}
\end{equation}
For the ground state $(0,0)$, $(1,0)$, and $(1,1)$, the marginal purity of two coupled oscillators can be expressed as
\begin{align}
    \mathcal{P}^{(0,0)}&=\bigg(1+\frac{\mathcal{Q} \mu _{\theta }^2}{\left(\mu _{\theta }^2+1\right)^2}\bigg)^{-\frac{1}{2}}, \label{eq60}
\end{align}
\begin{widetext}
\begin{align}
    \mathcal{P}^{(1,0)}&=\frac{\left(\mu _{\theta }^2+1\right) }{4}\frac{\left(4 \mu _{\theta }^8+4 (\mathcal{Q}+2) \mu _{\theta }^6+(3 \mathcal{Q} (\mathcal{Q}+4)+8) \mu _{\theta }^4+4 (\mathcal{Q}+2) \mu _{\theta }^2+4\right)}{ \left(\mu _{\theta }^4+(\mathcal{Q}+2) \mu _{\theta }^2+1\right)^{5/2}}\\
    \mathcal{P}^{(1,1)}&=\frac{\mu _{\theta }^2+1}{ \left(\mu _{\theta }^4+(Q+2) \mu _{\theta }^2+1\right){}^{9/2}} \bigg[1+ \mu _{\theta }^{16}+4 \mu _{\theta }^{12}+32 \mu _{\theta }^{10}+54 \mu _{\theta }^8+32 \mu _{\theta }^6+4 \mu _{\theta }^4+\frac{9}{16} \mathcal{Q}^4 \mu _{\theta }^8\\
    &\qquad+\mathcal{Q}^3 \left( \mu _{\theta }^{10}+\frac{9}{2} \mu _{\theta }^8+ \mu _{\theta }^6\right)+6\mathcal{Q}^2 \left( \mu _{\theta }^{12}+ \mu _{\theta }^{10}+\frac{3}{2} \mu _{\theta }^8+\mu _{\theta }^6+ \mu _{\theta }^4\right)+24\mathcal{Q} \left( \mu _{\theta }^{12}+ \mu _{\theta }^{10}+ \mu _{\theta }^6+ \mu _{\theta }^4\right) \bigg]\notag
\end{align}
\end{widetext}

At initial time $(t_{0})$ in Eq.~(\ref{eq29}), and where $M(t_0)$ is equal 1. The equation (\ref{eq65}) becomes \\
\begin{equation}
    \mathcal{Q}(t_0)=\frac{\left(\vartheta_1-\vartheta _2\right)^2}{\vartheta_1 \vartheta_2}\label{eq71}
\end{equation}
The findings align with those published in the literature \cite{1stchapter 9}. In Eq. (\ref{eq60}), it is clear that when the mixing angle $\mu_\theta=0$, the purity becomes $\mathcal{P}^{(0,0)}=1$. Also, when $\mu_{\theta}=1$ (which is $\theta=\frac{\pi}{4}$), the purity takes the formula 
\begin{equation}
     \mathcal{P}^{(0,0)}\bigg(\theta=\frac{\pi}{4}\bigg)=\left(\frac{4}{\mathcal{Q}(t_0)+4}\right)^{\frac{1}{2}}
\end{equation}
The results obtained here are in line with those presented in the literature \cite{1stchapter 4,1stchapter 10prime}.\\
It should be emphasized that a formal analogy can be drawn between the approach followed here using the Lewis-Riesenfeld invariant and the Wigner function formulation, and the approach used by Makarov \cite{Intro15} in terms of the Schmidt decomposition in the time-independent scenario  in case of weak coupling regim $\vartheta_1 \sim \vartheta_2 \sim\omega_1\sim\omega_2$ (where $\mathcal{Q}(t_0)=0$). In the static regime of the model under consideration, when all parameters remain time-independent, the link between both approaches can be established, and the linear entropy $S_L^{(n,m)} = 1 - \mathcal{P}^{(n,m)}$ obtained from our approach is connected to the Schmidt parameter $K$ introduced by Makarov as:
\begin{equation}
   S_L=1-K^{-1}=1-\sum\limits_{k=0}^{n+m}\lambda_k^2
\end{equation}
or equivalently:
\begin{widetext}
\begin{align}
 \mathcal{P}(n,m)&=\frac{(\mu _{\theta }^2+1)}{(n!m!)^{2}}
 \frac{d^n}{du^n} \frac{d^m}{dv^m}\frac{d^n}{da^n} \frac{d^m}{db^m}\bigg(-\frac{1}{\mathcal{A}+\mathcal{B}}\bigg)\Bigg|_{u,v,a,c=0}\notag\\
&=\frac{(n!m!)^2}{(1+\mu_\theta^2)^{2(m+n)}}\sum\limits_{k=0}^{n+m}\left[\frac{\mu_\theta^{2(k+n)}}{k!(m+n-k)!}\left(P_{n}^{(-(1+m+n),m-k)}\left(-\frac{2+\mu_\theta^2}{\mu_\theta^2}\right)\right)^2\right]^2
\end{align}
\end{widetext}
Here, $\mathcal{A}$ and $\mathcal{B}$ are defined in Eq.(\ref{eq58}), and $\lambda_k$ represents the Schmidt modes. The relationship above makes the explicit link between the two approaches clear, since it is apparent that the bipartite entanglement parameters we have been analyzing in terms of the purity $ \mathcal{P}^{(n,m)} $ and the Schmidt parameter $K$ by Makarov carry the same amount of information regarding the degree of entanglement of the system from their different viewpoints, the former using the Wigner function, while the latter relies on the Schmidt decomposition of the stationary state.

\section{Results and Discussion \label{sec5}}
Our work is more general because it covers all states $(n,m)$, not only the vacuum state. We have illustrated the entanglement of different states $(n,m)$, where the two coupled harmonic oscillators have the unit mass and the frequency is changed linearly. We have set the parameters in the Eqs. (\ref{equation15}) and (\ref{eq16}) $\left( \eta_i=\gamma_i=1,\, \text{and} \,\,\beta_i=0 \quad \forall i\in (1,2)\right)$\cite{1stchapter 3,1stchapter 4,1stchapter 12}. So the linearly independent solutions of the classical equation of motion $(\mathcal{F}_i(t),\, \text{and} \,\mathcal{G}_i(t) \quad \forall i\in (1,2))$ are written as 
\begin{align}
\mathcal{F}_i(t) =
\begin{cases}
\mathcal{F}_{i,\text{init}}(t) = \cos(\vartheta_{i,\text{init}} t), & t < 0 \\[6pt]
\mathcal{F}_{i,\text{fin}}(t) = \cos(\vartheta _{i,\text{fin}} t), & t > 0
\end{cases}
\end{align}
and
\begin{align}
\mathcal{G}_i(t) =
\begin{cases}
\mathcal{G}_{i,\text{init}}(t) = \sin(\vartheta_{i,\text{init}} t), & t < 0 \\[6pt]
\mathcal{G}_{i,\text{fin}}(t) =\left(\frac{\vartheta _{i,\text{init}} t}{\vartheta _{i,\text{fin}} t}\right) \sin(\vartheta _{i,\text{fin}} t), & t > 0
\end{cases}\label{eq69}
\end{align}
\begin{widetext}
where
\begin{equation}
\vartheta_{i}(t)=
\begin{cases}
  \vartheta_{i,\text{init}}=\vartheta_{i},   & t < 0 \\[6pt]
   \vartheta_{i,\text{fin}}=\vartheta_{i}\left( 1+\beta_0\right)^\frac{1}{2},  & t > 0
\end{cases}
\end{equation}
The label (init) designates the initial time, and (fin) designates the late time.\\
And $\mathbb{g}_{-i}(t)$, $g_{+i}(t)$ and $\mathbb{g}_{0i}(t)$ are defined as
\begin{align}
    \mathbb{g}_{-i}(t)&=
\begin{cases}
  \mathbb{g}_{-i,\text{init}}=1,& t < 0 \\[6pt]
  \mathbb{g}_{-i,\text{fin}}=\left(\frac{\vartheta _{i, \text{init}}}{\vartheta _{i, \text{fin}}}\right)^2+\left[1-\left(\frac{\vartheta _{i, \text{init}}}{\vartheta _{i, \text{fin}}}\right)^2\right] \cos ^2\left( \vartheta _{i, \text{fin}} t\right),& t > 0
\end{cases}\\
\mathbb{g}_{+i}(t)&=
\begin{cases}
  \mathbb{g}_{+i,\text{init}}=\vartheta _{i, \text{init}}^2,   & t < 0 \\[6pt]
  \mathbb{g}_{+i,\text{fin}}=\vartheta _{i, \text{fin}}^2-\left(\vartheta _{i,\text{fin}}^2-\vartheta _{i,\text{init}}^2\right) \cos ^2\left(\vartheta _{i,\text{fin}} t\right),  & t > 0
\end{cases}\\
\mathbb{g}_{0i}(t)&=
\begin{cases}
  \mathbb{g}_{0i,\text{init}}=0,   & t < 0 \\[6pt]
  \mathbb{g}_{0i,\text{fin}}=\left(\frac{\vartheta _{i,\text{fin}}^2-\vartheta _{i,\text{init}}^2}{\vartheta _{i,\text{fin}}}\right) \sin \left( \vartheta _{i,\text{fin}}t\right) \cos \left(\vartheta _{i,\text{fin}}t\right),  & t > 0
\end{cases}
\end{align}
\end{widetext}
and the frequency $\Omega_i$ in Eq. (\ref{eq20}) reduce to 
\begin{equation}
     \Omega_i(t)=
\begin{cases}
 \Omega_{i,\text{init}}=\vartheta _{i,\text{init}},   & t < 0 \\[6pt]
\Omega_{i,\text{fin}}=\vartheta _{i,\text{init}},  & t > 0
\end{cases}\label{eq74}
\end{equation}
For $( t < 0 )$, the parameter $\mathcal{Q}$ in Eq. (\ref{eq65}) is given by 
\begin{equation}
    \mathcal{Q}(t < 0) = \frac{(\vartheta_1 - \vartheta_2)^2}{\vartheta_1 \vartheta_2} \label{eq79}
\end{equation}
As a result of Eqs. (\ref{eq71}) and (\ref{eq79}), we get the following expression 
\begin{equation}
    \mathcal{Q}(t \leq 0) = \frac{(\vartheta_1 - \vartheta_2)^2}{\vartheta_1 \vartheta_2} \label{eq79}
\end{equation}
Consequently, in the case of $(t \leq 0)$, the purity corresponds to that of the time-independent coupled harmonic oscillators, as defined in the literature \cite{1stchapter 9,1stchapter 11}. In this scenario, the purity of the system depends solely on the quantum excitation numbers $(n,m)$, the mixing angle $\theta$, and the frequencies $(\vartheta_1,\vartheta_2)$. Therefore, the purity for $(t\leq 0)$ remains constant over time for $(t\leq 0)$. On the other hand, our focus here will be just on the case where $(t>0)$.\\

\subsection{ Impact of $\theta$ and $\vartheta_{2}$ on the linear entropy of the system.}
\begin{figure}[htbphtbp]
    \centering
    \includegraphics[width=8.5cm,height=6cm]{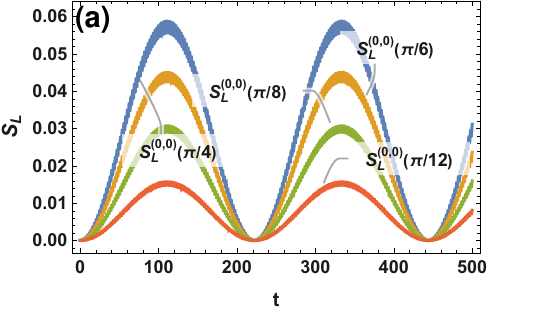}
    \includegraphics[width=8.5cm,height=6cm]{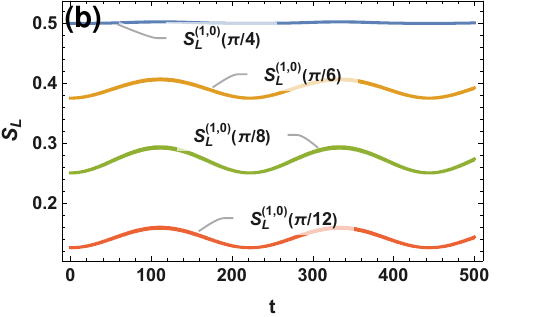}
    \includegraphics[width=8.5cm,height=6cm]{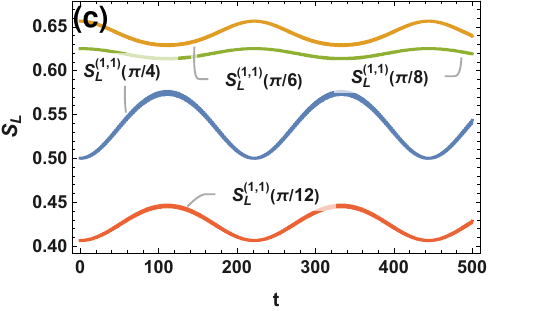}
    \caption{Evolution of linear entropy versus mixing angle $\theta$ for different quantum states $(n,m)$ with  $\vartheta_1=1$ and $\vartheta_2=1.01$}\label{fig1}
\end{figure}

To illustrate how the mixing angle $\theta$ influences the entropy of the system over time $t\in[0,500]$, we set $\vartheta_2=1.01$, $\vartheta_1=1$, and $\beta_0=1$. The given diagrams in Fig.~\ref{fig1} exhibit the time evolution of linear entropy $S_L$ for various states of the quantum systems: $(0,0)$, $(1,0)$, and $(1,1)$ for low frequency $\vartheta_2 = 1.01$. Under nearly resonant conditions ($\vartheta_1\simeq\vartheta_2$), all three panels (a)-(c) occur as slow oscillations of linear entropy across four different values of $\theta$. The amplitude of these oscillations varies depending on the quantum system state $(n, m)$; higher-order systems have stronger oscillations, influenced by the mixing angle $\theta$. The entropy reaches its maximum value for $\theta=\frac{\pi}{4}$. The multiple peaks observed in panels (a)-(c) show that all states exhibit quasi-periodic behavior. And the higher excitations levels exhibit a higher entanglement.

\begin{figure}[htbphtbp]
    \centering 
    \includegraphics[width=8.5cm,height=6cm]{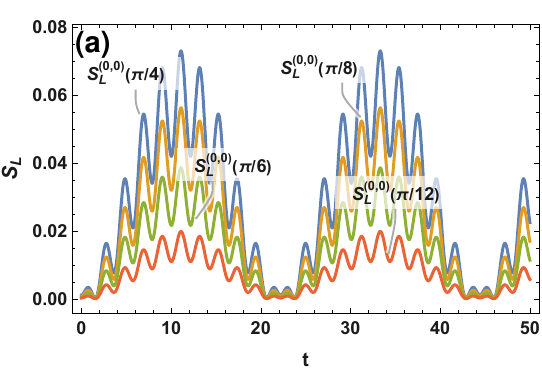}
    \includegraphics[width=8.5cm,height=6cm]{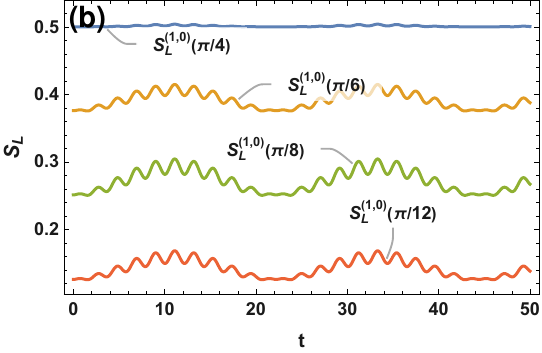}
    \includegraphics[width=9.6cm,height=6cm]{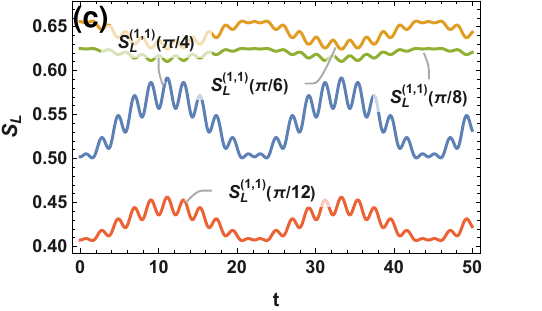}
    \caption{Evolution of linear entropy versus mixing angle $\theta$ for different quantum states $(n,m)$ with $\vartheta_1=1$ and $\vartheta_2=1.1$}\label{fig2}
\end{figure}

\begin{figure}[htbphtbp]
    \centering
    \includegraphics[width=8.5cm,height=6cm]{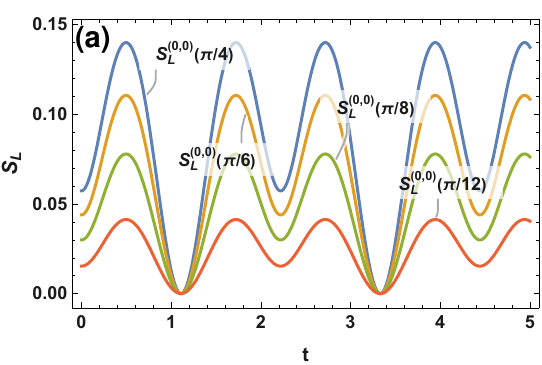}
    \includegraphics[width=8.5cm,height=6cm]{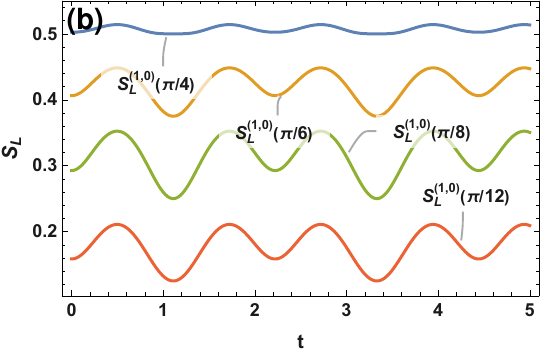}
    \includegraphics[width=8.5cm,height=6cm]{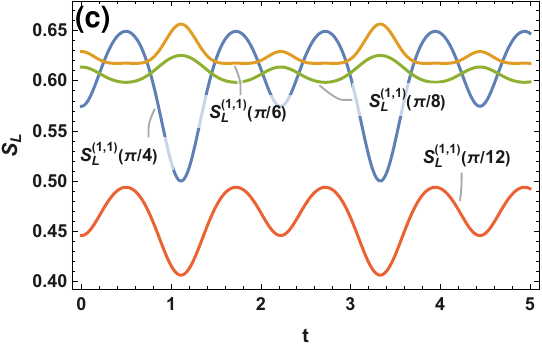}
    \caption{Evolution of linear entropy versus mixing angle $\theta$ for different quantum states $(n,m)$ with $\vartheta_1=1$ and $\vartheta_2=2$.}\label{fig3}
\end{figure}

When $\vartheta_2$ is increased to $ 1.1$, one observes highly improved oscillatory dynamics within a much smaller timescale $(t \leq 50)$, as illustrated in Fig.~\ref{fig2}. The fast, high-frequency oscillations seen on the left side of the figure show an increase in amplitude modulation. They demonstrate beating phenomena caused by an increase in frequency of $\vartheta_2$. The graphs seen on the middle and right side depict less regular oscillations with entropy fluctuations that seem to experience damping or modulation with time. Quantum states show different degrees of sensitivity to frequency fluctuations depending on the mixing angle $\theta$, and excitation levels $(n,m)$.
Fig.~\ref{fig3} presents the most interesting dynamical behavior of all the presented figures. By increasing the value of the parameter $\vartheta_2$ to 2, and examining evolution on a small timescale ($t \leq 5$), there are clear periodic oscillations of linear entropy in all three subgraphs (a)-(c). The regular oscillatory patterns across different quantum states $(n,m)$ indicate that strong detuning produces highly coherent and well-defined dynamics. The high-frequency oscillations reflect the enhanced frequency mismatch between system components, resulting in pronounced periodic modulation of the entropy evolution without significant damping or complexity.
Indeed, these parameters, $\theta$ and $\vartheta_2$, play a significant role in establishing how the changes in the linear entropy of the quantum system depend on time. An increment of the value of $\vartheta_2$ from 1.01 to 2 causes considerable changes in the behavior of linear entropy of  the system. First, when $\vartheta_2$ is relatively small, that is, when $\vartheta_2=1.01$, the variation in the value of entropy demonstrates slow oscillations within a certain period of time; in contrast, if $\vartheta_2$ is large, that is, when $\vartheta_2=2$, then the variation in the value of entropy is characterized by fast oscillations occurring during short time periods ($t \leq 5$). Thus, it was proved that the increment of the value of $\vartheta_2$ leads to the transition from slow chaos-like variations in entropy to the fast oscillatory behavior of its values. In addition, the value of $\theta$ affects the dependence of changes in quantum states $(n,m)$, controlling the amplitude of oscillations and their average values. In general, $\theta$ and $\vartheta_2$ form a control system, in which the value of $\vartheta_2$ regulates the timescale of dynamics, and $\theta$ affects the values of particular states.
\subsection{ Sensitivity of the Linear Entropy to $\beta_0$}
\begin{figure}[htbphtbp]
    \centering
    \includegraphics[width=8.5cm,height=6cm]{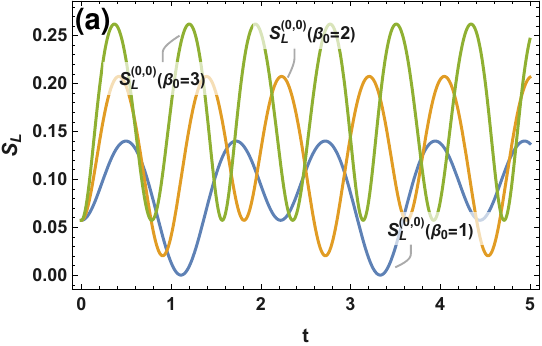}
    \includegraphics[width=8.5cm,height=6cm]{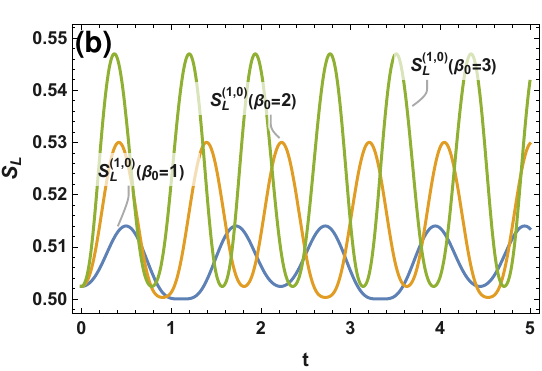}
    \includegraphics[width=8.5cm,height=6cm]{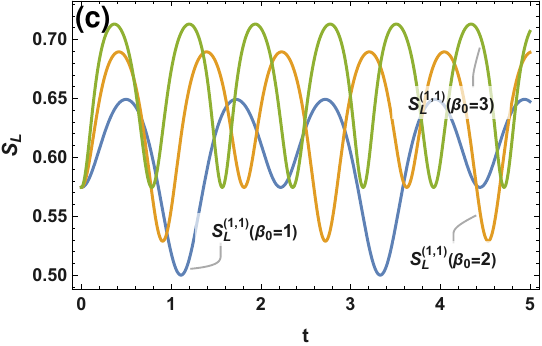}
    \caption{Evolution of linear entropy versus $\beta_0$ for several states $(n,m)$ with $\theta=\frac{\pi}{4}$, $\vartheta_1=1$, and $\vartheta_2=2$.}\label{fig4}
\end{figure}

The impact of the parameter $\beta_0$ on the linear entropy dynamics of several quantum states $(n,m)$ under constant parameters ($\vartheta_1=1, \vartheta_2=2$, and $\theta=\frac{\pi}{4}$) is shown in Fig~\ref{fig4}. In the three plots, the evolution of linear entropy $S_L^{(n,m)}$ within a brief period of time ($t \leq 5$) with the variation of $\beta_0$ for different quantum states $(n,m)$. These results imply that the parameter $\beta_0$ serves as an extremely sensitive control parameter since the modification of its value causes significant changes in terms of the amplitudes and the mean values of oscillations of entropy for each individual quantum state considered. Hence, the linear entropy is highly sensitive to changes in the value of $\beta_0$. Each of the considered quantum states $(n,m)$ has a different degree of sensitivity to this parameter since different quantum states demonstrate different amplitudes of oscillation under varied $\beta_0$. However, the oscillating dynamics is stable in all three panels (a)-(c) irrespective of changes in $\beta_0$. Consequently, it could be argued that the oscillating dynamics is stable, while $\beta_0$ affects the oscillations' properties.
\subsection{ Dependence of the linear entropy on coupling strength $\epsilon$}
\begin{figure}[htbphtbp]
    \centering
    \includegraphics[width=8.5cm,height=6cm]{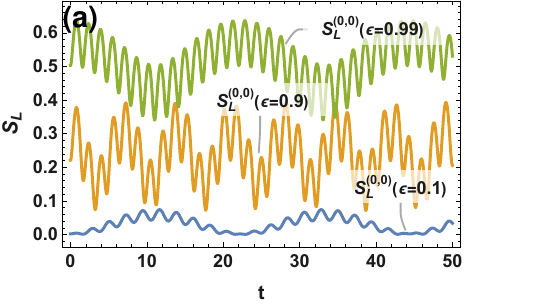}
    \includegraphics[width=8.5cm,height=6cm]{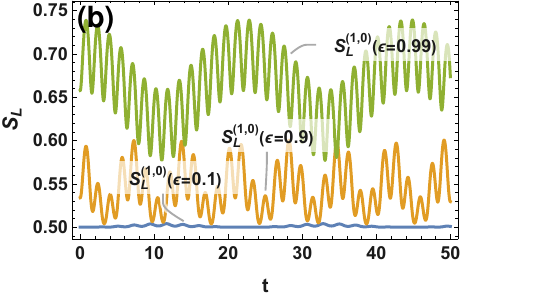}
    \includegraphics[width=8.5cm,height=6cm]{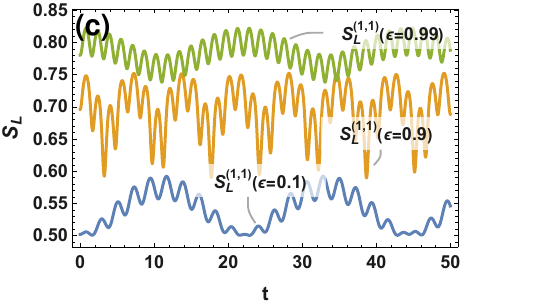}
    \caption{Evolution of linear entropy versus $\epsilon$ for several states $(n,m)$ with $\theta=\frac{\pi}{4}$, $\omega_1=\omega_2=1$, and $\beta_0=1$.}\label{fig5}
\end{figure}

Fig.~\ref{fig5} shows the influence of the coupling strength parameter $\epsilon$ on the time development of the linear entropy evolution of several quantum states $(n,m)$ under the fixed set of parameters ($\omega_1 = 1, \omega_2 = 1, \theta =\frac{\pi}{4}$, and $\beta_0 = 1$). Three panels (a)-(c) represent the evolution of linear entropy $S_L$ over a long period of time ($t \in [0, 50]$) for different quantum states having different values of the coupling constant ($\epsilon = 0.1, \epsilon = 0.5$, and $\epsilon = 0.99$). It should be noted that the coupling constant plays a crucial role in the process under consideration. In fact, the analysis shows that the coupling constant $\epsilon$ is one of the key parameters that determine the dynamics of the studied system: small values of the coupling constant—weak coupling—($\epsilon  = 0.1$) result in lower amplitude entropy oscillation and smaller mean values, while large values of the coupling constant—strong coupling—($\epsilon  = 0.99$) produce high amplitude oscillations and relatively high mean entropy values. Besides, different quantum states are characterized by different responses to changes in the value of $\epsilon$.

\subsection{Strong coupling regime of resonance case for different states}
\begin{figure}[htbphtbp]
    \centering
    \includegraphics[width=8.5cm,height=6cm]{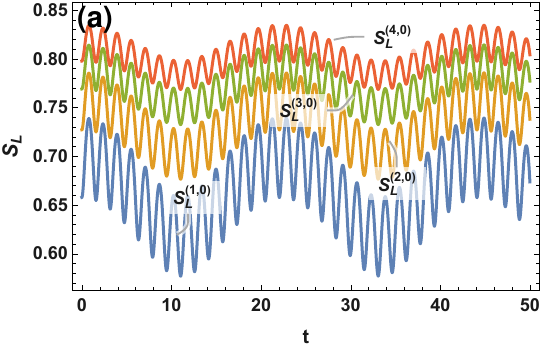}
    \includegraphics[width=8.5cm,height=6cm]{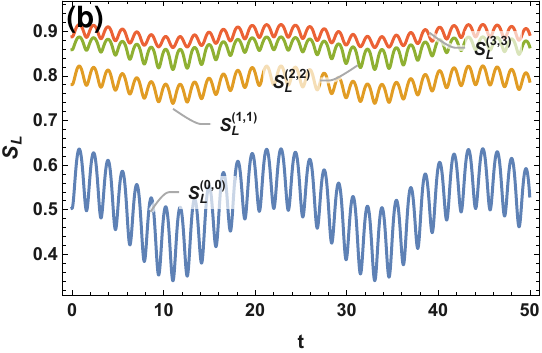}
    \includegraphics[width=8.5cm,height=6cm]{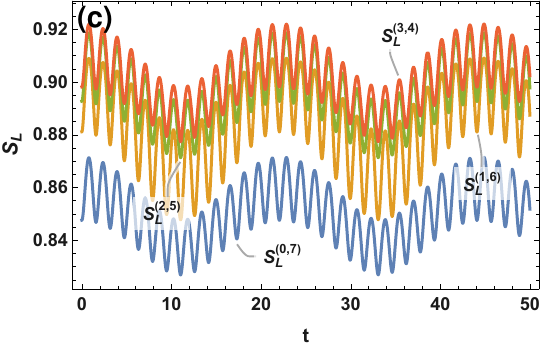}
    \caption{Evolution of linear entropy for different quantum states $(n,m)$ with $\theta=\frac{\pi}{4}$, $\omega_1=\omega_2=1$, and $\beta_0=1$}\label{fig6}
\end{figure}

Fig.~\ref{fig6} presents the dynamics of the linear entropy $S_L$ for quantum states $(n,m)$ in the strong coupling resonance case ($\omega_1 =\omega_2 = 1, \theta = \frac{\pi}{4}, \beta_0= 1$ and $\epsilon = 0.99$). Three plots (a)-(c) illustrate the time dependence of the linear entropy $S_L$ during the period $t \in [0, 50]$. For the strong coupling resonance regime, all considered states possess very organized and stable oscillatory behavior with clear periodicity, e.g., $((3,3), (3,4), (2,5))$. Different quantum states have different entropy intervals: entropy oscillations occur strongly for low-order quantum states $((0,0), (1,1))$. All oscillations of all quantum states are perfectly synchronous and damped. This means that the strong coupling resonance case provides very good dynamics. The oscillation amplitudes and mean values of the entropy are characteristic for each particular quantum state, which means that different quantum states behave differently when interacting with the system in the strong coupling resonance regime.  

\begin{comment}
\end{comment}

\section{Conclusion\label{sec6}}
In this study, the evolution of quantum entanglement in two coupled harmonic oscillators under the influence of a time-varying coupling is explored using the Lewis-Riesenfeld approach, together with a formulation in the Wigner phase space. Using this approach, exact forms of the wave functions, as well as analytic forms of purity and linear entropy measures of entanglement were formulated for any pair of excitation numbers $(n,m)$ in a completely nonperturbative and nonadiabatic way.\par
Our analysis revealed that the dynamics of the entanglement, quantified by the linear entropy $S_L=1-\mathcal{P}$, is very sensitive to the values of the model parameters. The effect of the parameters of the phase differences ($\theta$, $\vartheta_2$) on the entanglement was demonstrated to be related to the oscillatory nature of its time evolution: increasing those parameters leads to a transition from a slow oscillating behavior to fast and stable periodical evolution, while $\theta$ regulates the state-dependent amplitudes. On the other hand, the frequency parameter $\beta_0$ enables an accurate adjustment of the oscillation features without changing the dynamic regime, while the dependence of the entanglement on the coupling strength $\epsilon$ became evident. Increasing $\epsilon$ systematically raises the amplitude and average value of $S_L(t)$. Indeed, most surprisingly, in the case of resonance when $\omega_1 = \omega_2 = 1$ accompanied by high values of the coupling strength parameter $\epsilon \approx 0.99$, a stable periodic evolution of the linear entropy is observed without any damping effect. In addition, perfect synchronization of the oscillations in the resonance regime does not lead to quantum coherence saturation; thus, an entangled state can persist forever.\par
These results are consistent with and extend the findings of Galve et al. \cite{Intro18}, Roque and Roversi \cite{Intro19}, and Bastidas et al. \cite{Intro14}, who established the central role of normal-mode stability in governing entanglement dynamics in time-dependent coupled oscillator systems. Our work enriches this picture by providing explicit analytical parameter control over the entanglement dynamics and by demonstrating that even modest tuning of the coupling and detuning parameters can qualitatively transform the dynamical behavior of quantum correlations.\par
In summary, all these results reveal a clear relationship between the physical characteristics of the time-dependent coupling and the entanglement dynamics of the system under consideration. which can be used for controlling quantum correlation in continuous variable systems and in quantum information processing tasks.

\end{document}